# Evaluation of S-Parameters Similarity with Modified Hausdorff Distance

Yuriy Shlepnev
*Simberian Inc., 615 Hampton Dr. Unit B306, Venice, CA 90291, USA*
shlepnev@simberian.com

*Abstract* — A new similarity measure for two sets of S-parameters is proposed. It is constructed with the modified Hausdorff distance applied to S-parameter points in 3D space with real, imaginary and normalized frequency axes. New S-parameters similarity measure facilitates automation of the analysis to measurement validation, comparison of models and measurements obtained with different tools, as well as finding similar S-parameter models or similar elements within S-matrices.

## I. INTRODUCTION

Bandwidth required for signal integrity analysis of PCB and packaging interconnects is growing with the increase of data rates. Evaluation of model accuracy requires validation with the measurements – this is a necessary element of successful design process. A systematic approach for the analysis to measurement validation was recently introduced in [1], [2]. Though, the last step in the process was a visual estimate of the closeness of models to measured data. A formal measure is needed for automation. Feature Selective Validation (FSV) method [3] can be used for such purpose. However, it is rather complicated, has too many parameters and can be applied only to amplitudes of S-parameters. A single number similarity measure for S-parameters is introduced in this paper and illustrated with practical examples.

## II. S-PARAMETERS SIMILARITY MEASURE

Let's consider two S-parameter sets *SA* and *SB* defined as follows:

$$SA = \{SA(fa_k), k=1,...,K\}, \quad SA(fa_k) \in C^{N \times N}$$
$$SB = \{SB(fb_m), m=1,...,M\}, \quad SB(fb_m) \in C^{N \times N}$$

$SA(fa_k)$ and $SB(fb_k)$ are N by N complex matrices with the elements defined at each frequency point as $sa_{i,j}(fa_k), k=1,...,K$ and $sb_{i,j}(fb_m), m=1,...,M$ with $i,j = 1,...N$. For simplicity, the matrix element indexes *i,j* are omitted in some expressions below.

If two sets have exactly the same number of collocated frequency points $K = M$ and $f_k = fa_k = fb_k$, each element of S-matrix can be treated as a complex vector with dimension equal to the number of frequency points *K* and distance between two elements of S-matrices as follows:

$$d_{abs}(sa, sb) = \frac{1}{K} \sum_{k=1}^{K} |sa(f_k) - sb(f_k)| \quad (1)$$

$$d_{rms}(sa, sb) = \sqrt{\frac{1}{K} \sum_{k=1}^{K} |sa(f_k) - sb(f_k)|^2} \quad (2)$$

Where $|\cdot|$ denotes amplitude of the complex vectors or Euclidian vector length in 2D real-imaginary space. Such distances are often used as error metrics for optimization and for model convergence evaluation. There are other possibilities for distances defined with vector space norms. Unfortunately, distances like $d_{abs}$ and $d_{rms}$ may be not useful for comparison of simulations with the measurements during the validation process. A model and measured S-parameters may be sampled at different frequency points over different bandwidth. Interpolation can be used, but it may introduce additional errors. Distances like (1) and (2) are also not useful in cases if only similarity of the S-parameter sets has to be evaluated – cases with slightly shifted resonances for instance.

To compare two sets of S-parameter with possible different frequency sampling, any directed distance between two point sets defined in [4] can be used. Similar distances can be defined between two sets of S-parameters in 3D space formed by real, imaginary and normalized frequency axes. First, we convert each element of *SA* and *SB* sets into points with 3 coordinates as follows:

$$sa^k = \left(\text{Re}(sa(fa_k)), \text{Im}(sa(fa_k)), fa_k/f_{norm}\right)$$
$$sb^m = \left(\text{Re}(sb(fb_m)), \text{Im}(sb(fb_m)), fb_m/f_{norm}\right)$$

$f_{norm}$ is normalization frequency – it defines unit along the Z-coordinate. This is a plot in real-imaginary-frequency (RIF) space. Its projection into XY-plane is just a regular polar plot. It can be called 3D spiral plot (causal S-parameters are always spiral-like with clockwise rotation with increase of frequency). Note that values of S-parameters are bounded by unit for passive systems such as interconnects.

A distance between a point $sa^k$ and a set of points $sb = \{sb^m, m=1,...,M\}$ can be defined as follows:

$$d_{rif}(sa^k, sb) = \min_{m=1,...,M} |sa^k - sb^m| \quad (3)$$

Where $|\cdot|$ is regular Euclidian norm or length of the vectors, but now in 3D RIF space (unlike (1) and (2)). Maximal value of (3) over all points $sa^k$ is the regular Hausdorff distance. It would be too pessimistic to use it for similarity evaluation.

Modified Hausdorff distance [4] between two elements of S-matrices is defined as follows:

$$d_{MH}(sa, sb) = \frac{1}{K}\sum_{k=1}^{K} d_{rif}(sa^k, sb) \quad (4)$$

This distance is computed separately for each element of S-matrix (matrix indexes $i,j$ are not shown for simplicity). The distance (4) is not symmetric in general, but can be converted into such by defining it as $\max(d_{MH}(sa,sb), d_{MH}(sb,sa))$ [4]. Here we assume that the set $SA$ may have smaller number of frequency points comparing to set $SB$. The one-directional distance is more suitable for such cases. Also, only subsets of frequency points over the same frequency bandwidth are used for computations. Multiple other choices for the distance are possible [4], but (4) is selected as the best for the object matching (objects in our case are S-parameter element points in 3D RIF space). Also, it is easy to show that with the equidistant collocated sampling and with $f_{norm} = 1$ (no normalization) $d_{MH}(sa, sb) = d_{abs}(sa, sb)$.

Distance between two sets of S-parameters $SA$ and $SB$ can be defined as follows:

$$d_{MH}(SA, SB) = \max\left[d_{MH}(sa_{i,j}, sb_{i,j}), i,j = 1,...,N\right] \quad (5)$$

Note that distances $d_{abs}$ or $d_{rms}$ can be also used with possible interpolation (no interpolation is required for $d_{MH}$). $d_{MH}$ can be directly used to evaluate the similarity, but more intuitive similarity measure for S-matrix elements and for the whole matrix can be defined as follows (similar to the quality measures introduced in [5]):

$$SPS(sa, sb) = 100 \cdot \max(1 - d_{MH}(sa, sb), 0)\% \quad (6)$$

$$SPS(SA, SB) = \min\left(SPS(sa_{i,j}, sb_{i,j}), i,j = 1,...,N\right) \quad (7)$$

Other distances (1) and (2) can be also used to compute SPS. SPS measure is bounded by 0 for cases with no similarity at all and 100% for exactly the same sets of data (identical data sets). The other tiers or levels of similarity can be introduced for a particular set of problems as demonstrated in the next section.

III. ILLUSTRATIVE AND PRACTICAL EXAMPLES

As the first example let's apply the algorithm to analysis to measurement validation for a simple 2-inch strip line segment with two launches and connectors from CMP-28 validation platform from Wild River Technology (WRT) [1]. Magnitude of simulated and measured transmission and reflection parameters are plotted in Fig. 1. Simulation was done with de-compositional analysis [5] in Simbeor software and measurements are provided by WRT. The correlation looks good up to 30-35 GHz. Though, to make a definite conclusion, we need to compare the phases of the corresponding S-parameters as well. Instead, let's take a look at the same S-parameter in the RIF space - 3D spiral plot is shown in Fig. 2.

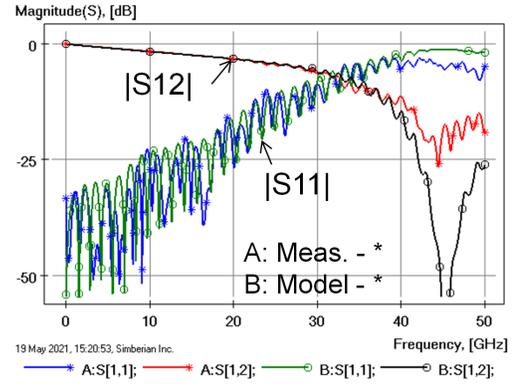

Fig. 1. Simulation vs. measurement for 2-inch segment.

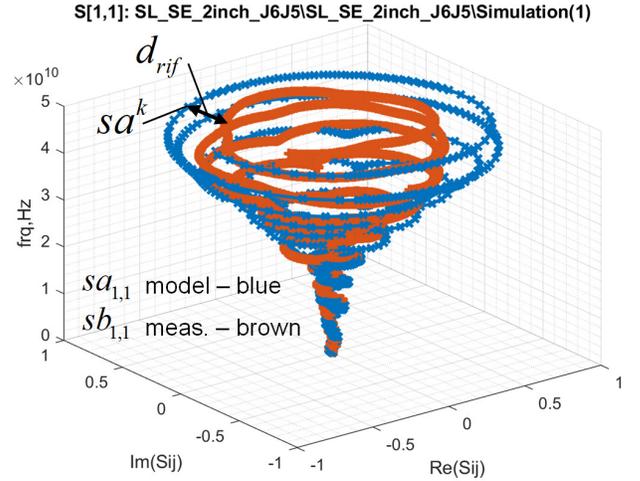

Fig. 2. 3D spiral plot for S11 parameters for 2-inch segment.

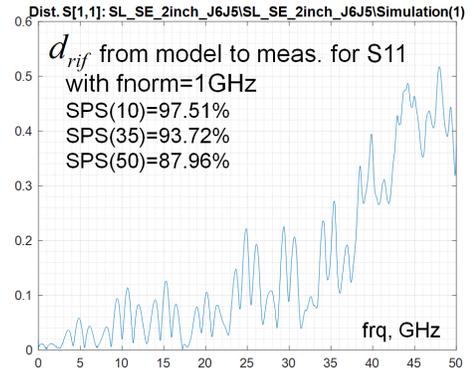

Fig. 3. Distance for S11 parameters for 2-inch segment.

For illustrative purpose, a frequency point $sa^k$ and the distance $d_{rif}(sa^k, sb)$ defined by (3) are also shown on the plot in Fig. 2. Technically, it is the minimal distance from one of the blue points to a point on the brown curve. It can be also the smallest distance directly to the brown curve, if interpolation is allowed. Note that the corresponding point on the brown curve $sb^m$ may be not necessarily at the same frequency as $sa^k$. If the sampling is collocated and the curves are very close, two points are at the same frequency, if the normalization frequency is sufficiently small. However, if there are two resonances at slightly different frequencies (two loops shifted along the frequency axis), the distance between two different frequency points on those loops may be smaller than the distance at the same frequency point with sufficiently large normalization frequency. It allows comparison of S-parameters with similar features such as sharp resonances.

Dependency of the distance $d_{rif}(sa^k, sb)$ from the points in $sa_{1,1}$ (model) to $sb_{1,1}$ (measured) versus frequency is plotted in Fig. 3. We can see that the distance is growing with the frequency. SPS values (6) computed with bandwidth 10 GHz (SPS(10)) 35 GHz (SPS(35)) and 50 GHz (SPS(50)) are also shown in Fig. 3 – we can see that the similarity degrades with the increase of bandwidth.

The spiral plots for simulated and measured S12 parameters are shown in Fig. 4 and corresponding dependency of the distance from frequency in Fig. 5.

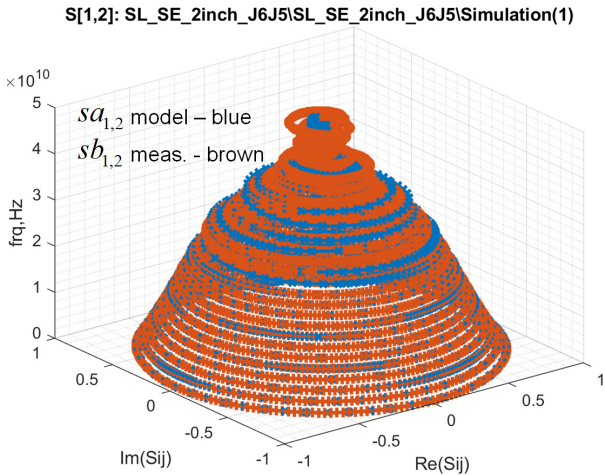

*Fig. 4. 3D spiral plot for S12 parameters for 2-inch segment.*

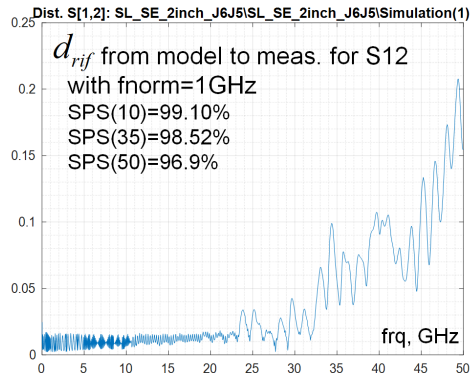

*Fig. 5. Distance for S12 parameters for 2-inch segment.*

Note that the SPS measure will depend on the normalization frequency. It basically defines the vicinity of each point along the frequency axis. To avoid sensitivity to sampling, the normalization frequency should be greater than the step frequency in set *SB* (measured data with equidistant frequency sweep). In this case, the frequency step in measured data was 10 MHz and selection of the normalization frequency within 100 MHz to 1 GHz produces practically the same results. 1 GHz normalization frequency was used for all examples here.

Finally, values of SPS computed for all test structures on CMP-28 validation platform [1] are shown in the table below for 3 different comparison bandwidths – 10 GHz, 35 GHz and 50 GHz.

| Model | Meas | SPS_SE 10 GHz | SPS_SE 35 GHz | SPS_SE 50 GHz |
|---|---|---|---|---|
| SL_SE_2inch_J6J5 | cmp28_strpl_2in_50ohm_p1J6_p2J5_s2p | 97.1513 | 92.5639 | 84.677 |
| SL_SE_8inch_J7J8 | cmp28_strpl_8inch_p1J7_p2J8_s2p | 97.8176 | 91.8262 | 80.9387 |
| SL_SE_Beatty_25Ohm_J28J27 | cmp28_strpl_Beatty_25ohm_p1J28_p2J27_s2p | 98.3164 | 91.7525 | 81.1544 |
| SL_SE_Resonator_J23J24 | cmp28_strpl_resonator_p1J23_p2J24_s2p | 98.5621 | 92.8552 | 82.7012 |
| SL_SE_Via_Capacitive_J18J17 | cmp28_strpl_via_capacitive_p1J18_p2J17_s2p | 94.9476 | 91.1739 | 82.8437 |
| SL_SE_Via_Backdrilled_J14J13 | cmp28_strpl_via_backdrilled_p1J14_p2J13_s2p | 97.1172 | 90.8311 | 82.0804 |
| SL_SE_2inch_Capacitive_J9J10 | cmp28_strpl_2in_Capacitive_p1J10_p2J09_s2p | 97.7805 | 93.0992 | 87.3275 |
| SL_SE_2inch_Inductive_J11_J12 | cmp28_strpl_2in_Inductive_p1J12_p2J11_s2p | 97.8352 | 93.8351 | 87.8757 |
| SL_DF_2inch | cmp28_strpl_diff_2inch_J39J40J35J36_s4p | 95.9985 | 91.087 | 83.0354 |
| SL_DF_6inch | cmp28_strpl_diff_6inch_J47J48J43J44_s4p | 96.8208 | 93.0776 | 85.1746 |
| MS_SE_2in_J1_J2 | cmp28_mstrp_2in_p1J1_p2J2 | 97.9111 | 94.7303 | 91.8845 |
| MS_SE_8in_J4_J3 | cmp28_mstrp_8inch_p1J4_p2J3 | 97.6372 | 95.3771 | 91.645 |
| MS_SE_Beatty_25Ohm_J25_J26 | cmp28_mstrp_Beatty_25ohm_p1J25_p2J26 | 96.5268 | 93.3182 | 89.9407 |
| MS_SE_Resonator_J21_J22 | cmp28_mstrp_resonator_p1J21_p2J22 | 98.0708 | 94.1929 | 90.5811 |
| MS_SE_GND_Voids_J74_J75 | cmp28_gnd_voids_p1J74_p2J75 | 97.6512 | 88.4187 | 83.5582 |
| MS_SE_GraduateCoplanar_J70_J69 | cmp28_graduate_coplanar_p1J70_p2J69 | 97.6924 | 94.4118 | 91.4621 |
| MS_SE_Via_Inductive_J15_J16 | cmp28_mstrp_via_inductive_p1J15_p2J16 | 96.6664 | 93.596 | 90.0153 |
| MS_SE_Via_Capasitive_J19_J20 | cmp28_mstrp_via_capacitive_p1J19_p2J20 | 96.5088 | 93.969 | 90.1057 |
| MS_SE_Via_Pathology_J65_J66 | cmp28_via_pathology_p1J65_p2J66 | 97.2525 | 91.9582 | 88.486 |
| MS_DF_2inch | cmp28_mstrp_diff_2inch_J38J37J34J33 | 95.4645 | 93.3429 | 90.407 |
| MS_DF_6inch | cmp28_mstrp_diff_6inch_J46J45J42J41 | 95.5751 | 93.9318 | 90.9123 |
| MS_DF_GND_Cutout | cmp28_mstrp_diff_gnd_cutout_J59J60J55J56 | 94.4506 | 91.4807 | 88.7113 |
| MS_DF_Vias | cmp28_mstrp_diff_vias_J49J50J51J52 | 95.6808 | 91.6811 | 88.4878 |

We can see that there is much better similarity at lower frequencies (10 GHz column) and it degrades with larger bandwidths (35 and 50 GHz columns). SPS was also computed for all structures on EvR-1 validation platform [2]. After additional visual inspection of data and explanations of the problems, the following tiers for the SPS values are suggested for the broadband interconnect problems: Good [99,100], Acceptable [90,99), Inconclusive [80,90), and Bad [0,80).

IV. CONCLUSION

A new S-parameters similarity measure is introduced in the paper on the base of modified Hausdorff distance applied to elements of S-matrices in 3D real-imaginary-frequency space (RIF space). Unlike norm-based similarity measures, it does not require identical frequency sampling or interpolation. The measure is simple and computationally straightforward. It is also intuitive – based on comparison of two sets of points in 3D RIF space, similar to image recognition. It is shown that tiers or levels can be introduced for a particular application domain. The approach satisfies 5 basic principles for automated validation method outlined in [3] and may compliment FSV technique.